\documentclass[prd,preprintnumbers,nofootinbib]{revtex4} 
\usepackage{graphicx} 
\usepackage{amsmath}
\usepackage{amsfonts,amsbsy}
\usepackage{amssymb}

\def\be{\begin{equation}}
\def\ee{\end{equation}}
\def\bea{\begin{eqnarray}}
\def\eea{\end{eqnarray}}

\def\fig#1{{Fig.~\ref{#1}}}

\def\beq{\begin{equation}}
\def\eeq{\end{equation}}
\def\bea{\begin{eqnarray}}
\def\eea{\end{eqnarray}}

\def\fig#1{{Fig.~\ref{#1}}}

\newcommand{\Lb}{\left(}
\newcommand{\Rb}{\right)}

\begin{document}

\title{\bf Charged particle multiplicities in pA interactions at the LHC 
from the Color Glass Condensate}

%\preprint{}

\author{Amir H. Rezaeian}
\affiliation{
Departamento de F\'\i sica, Universidad T\'ecnica
Federico Santa Mar\'\i a, Avda. Espa\~na 1680,
Casilla 110-V, Valparaiso, Chile }

\begin{abstract}
The forthcoming LHC measurement of hadron multiplicity in
proton-nucleus collisions is a crucial test of the $k_t$ factorization
and gluon saturation based models.  Here, we provide quantitative
predictions for the pseudorapidity distribution of charged particles
produced in minimum bias proton-nucleus collisions at the LHC based on
the idea of gluon saturation in the color-glass condensate
framework. Our formulation gives good descriptions of the LHC and
RHIC data for the charged-hadron multiplicities in both proton-proton
and nucleus-nucleus collisions, and also the deep inelastic scattering
at HERA at small Bjorken-x.

\end{abstract}

\maketitle

%---------------------------------------------------------------------------
\section{Introduction}

The Color Glass Condensate (CGC) formalism is a self-consistent
effective QCD theory at high energy (or small $x$) in which one
systematically re-sums quantum corrections which are enhanced by large
logarithms of $1/x$ and also incorporates non-linear high gluon
density effects which are important where the physics of gluon
saturation is dominant, for a review see Ref.~\cite{review} and
references therein.

In the CGC approach, the hadron production in proton-nucleus
collisions goes in two stages: production of gluons \cite{KTINC} and subsequently
the decay of gluon-jet or mini-jet into hadrons, namely hadronization
(or fragmentation in the standard pQCD language) \cite{ap}. The first stage of this process
is under theoretical control and can be calculated via the $k_t$
factorization. The jet decay, unfortunately, can be treated mostly
phenomenologically.  However, one may hope that the phenomenological
uncertainties would be reduced to few constants whose values will be
extracted from the experiment in other reactions and energy. Notice
that the inclusive gluon-production in the $k_t$ factorization has
been proven \cite{KTINC} at the leading log(1/x) approximation for
scatterings of a dilute partonic system on a dense one such as p+A
collisions at high energy\footnote{For the recent theoretical
development on this line including the effect of the running coupling
corrections to the $k_t$ factorization, see Ref.~\cite{kov1}.}. In
nucleus-nucleus collisions at around midrapidity, one has to deal with
a dense-dense scatterings, therefore the $k_t$ factorization and as a
consequence all saturation based predictions are less
reliable. Therefore, the upcoming data from p+A run at the LHC is very
crucial in order to test the $k_t$ factorization and discriminate
between various saturation models.

Already the first LHC data on hadron multiplicity in nucleus-nucleus
collisions call for a theoretical understanding of these data based on
QCD. One of the unexpected new feature of the LHC data has been
the very different power-law energy behavior of charged hadron
multiplicities in A+A compared to p+p collisions
\cite{Apb1,CMS}. Although this is still open problem, 
there have been some different approaches to accommodate this feature of
data within the saturation picture \cite{me1,JA}. We will show below that
the first-day experimental p+A data on charged hadron multiplicity can
also discriminate between these approaches.

In this letter, we provide quantitative predictions for the
pseudorapidity distribution of charged particles produced in minimum
bias proton-nucleus collisions at the LHC based on the $k_t$
factorization within the CGC framework. This approach has been also
successful to describe the charged hadron multiplicity in both p+p and
A+A collisions at the LHC \cite{me1,ap}, see also \cite{me3,raj,la}.
We refer the reader to Refs.~\cite{me1,ap,aa} for the technical
details. Here we shortly introduce the main formalism in Sec. II and then
show our results in Sec III.

\section{Main formulation}

The gluon jet production in A+B collisions can be described by $k_T$-factorization given by \cite{KTINC},
\begin{equation} \label{M1}
\frac{d \sigma}{d y \,d^2 p_{T}}=\frac{2\alpha_s}{C_F}\frac{1}{p^2_T}\int d^2 \vec k_{T} \phi^{G}_{A}\Lb x_1;\vec{k}_T\Rb \phi^{G}_{B}\Lb x_2;\vec{p}_T -\vec{k}_T\Rb,
\end{equation}
where $C_F=(N_c^2-1)/2N_c$ with $N_c$ being the number of colors,
$x_{1,2}=(p_T/\sqrt{s})e^{\pm y}$, $p_T$ and $y$ are the
transverse-momentum and rapidity of the produced gluon
jet. $\phi^{G}_{A}(x_i;\vec k_T)$ denotes the unintegrated gluon
density and is the probability to find a gluon that carries $x_i$
fraction of energy with $k_T$ transverse momentum in the projectile
A (or target B). The unintegrated gluon
density is related to the color dipole forward scattering amplitude,  
\beq \label{M2}
\phi^G_A\Lb x_i;\vec{k}_T\Rb=\frac{1}{\alpha_s} \frac{C_F}{(2 \pi)^3}\int d^2 \vec b d^2 \vec r_T
e^{i \vec{k}_T\cdot \vec{r}_T}\nabla^2_T N^G_A\Lb x_i; r_T; b\Rb,
\eeq
with notation
\beq \label{M3}
N^G_A\Lb x_i; r_T; b \Rb =2 N_A\Lb x_i; r_T; b \Rb - N^2_A\Lb x_i; r_T; b \Rb,
\eeq
where $r_T$ denotes the dipole transverse size and $\vec b$ is the
impact parameter of the scattering. For the value of strong-coupling
$\alpha_s$ we employ the running coupling prescription used in
Ref.~\cite{aa}.

The most important ingredient of the single inclusive hadron
production cross section which captures the saturation dynamics is the
fundamental (or adjoint) dipole cross section, the imaginary part of
the forward quark anti-quark scattering amplitude on a proton or
nucleus target $N_{A,p}\Lb x_i; r_T; b \Rb$.  In principle, the dipole cross-section can be
computed via the JIMWLK/BK evolution equations
\cite{jm,jm2}. For the recent progress in obtaining the impact-parameter dependent solution of 
the BK equation, see Ref.~\cite{bk-b}. It has been shown that the impact-parameter dependence of the
dipole amplitude is important for phenomenological studies \cite{JA,me1,ap,raj,aa}, including
for describing the diffractive data at HERA  \cite{b-cgc}. 
Following Refs.~\cite{me1,ap,aa} we use the b-CGC
saturation model \cite{b-cgc} which explicitly depends on
impact-parameter. This model approximately incorporates all known
properties of the exact solution to the BK equation including the
impact-parameter dependence of the scattering amplitude \cite{ap} and it also
describes the small-$x$ data at HERA \cite{b-cgc}.  For comparison,
we will confront the predictions coming from the b-CGC with the MCrcBK
\cite{JA}. The MCrcBK model effectively incorporates the
impact-parameter dependence into the BK equation solution via the
Monte Carlo implementation of the $k_t$ factorization by taking into
account fluctuations of the position of nucleon inside a nucleus
\cite{JA}. Notice that both the b-CGC and the MCrcBK model have been
quite successful to describe various experimental data in different
reactions. However, we will show here that these two models give
rather different predictions for the charged hadron multiplicity in p+A
collisions at the LHC.

The $k_t$ factorization has infrared divergence. By introducing a new
parameter $m_{jet}$ as mini-jet mass which mimics the
pre-hadronization effect, one can also regularize the
cross-section. Unfortunately, we do not know how mini-jet mass changes
with medium and kinematics. Therefore, one may model or chose the
value of mini-jet at lower energy and hope that it should be still
valid at higher energy for different rapidities and centralities. This
is the approach that all previous saturation based studies have
adopted, although, in different models, different values for $m_{jet}$
has been taken \cite{me1,JA,ap,aa,raj,kln}.  Therefore, the
uncertainties associated with our freedom in choosing the mini-jet
mass value should be considered and will be quantified here. In order
to take account of the difference between rapidity $y$ and the
measured pseudo-rapidity $\eta$, we employ the Jacobian transformation
between $y$ and $\eta$ \cite{ap}. The final-state hadronization
effects are incorporated via the Local Parton-Hadron Duality principle
\cite{me1,JA,ap,aa,raj}, namely the hadronization is a soft process
and cannot change the direction of the emitted radiation\footnote{One
should note that the main contribution of the $k_t$ factorization for
the multiplicity comes from small $p_T < 1.5$ GeV where the
fragmentation functions are not reliable. We recall that it is still
an open problem how to incorporate the fragmentation processes into
the CGC/saturation formalism.}.  This introduces only one extra
parameter which can be absorbed into the over-all normalization of the
cross-section. Therefore, we have only two unknown parameters in our
model, the overall normalization factor and the mini-jet mass which
are fixed at lower energy at midrapidity for central collisions. Then
our results at higher energies, different rapidities and centralities
can be considered as free-parameter predictions.

\section{DISCUSSION AND PREDICTIONS}

\begin{figure}[t]                            
                           \includegraphics[width=8 cm] {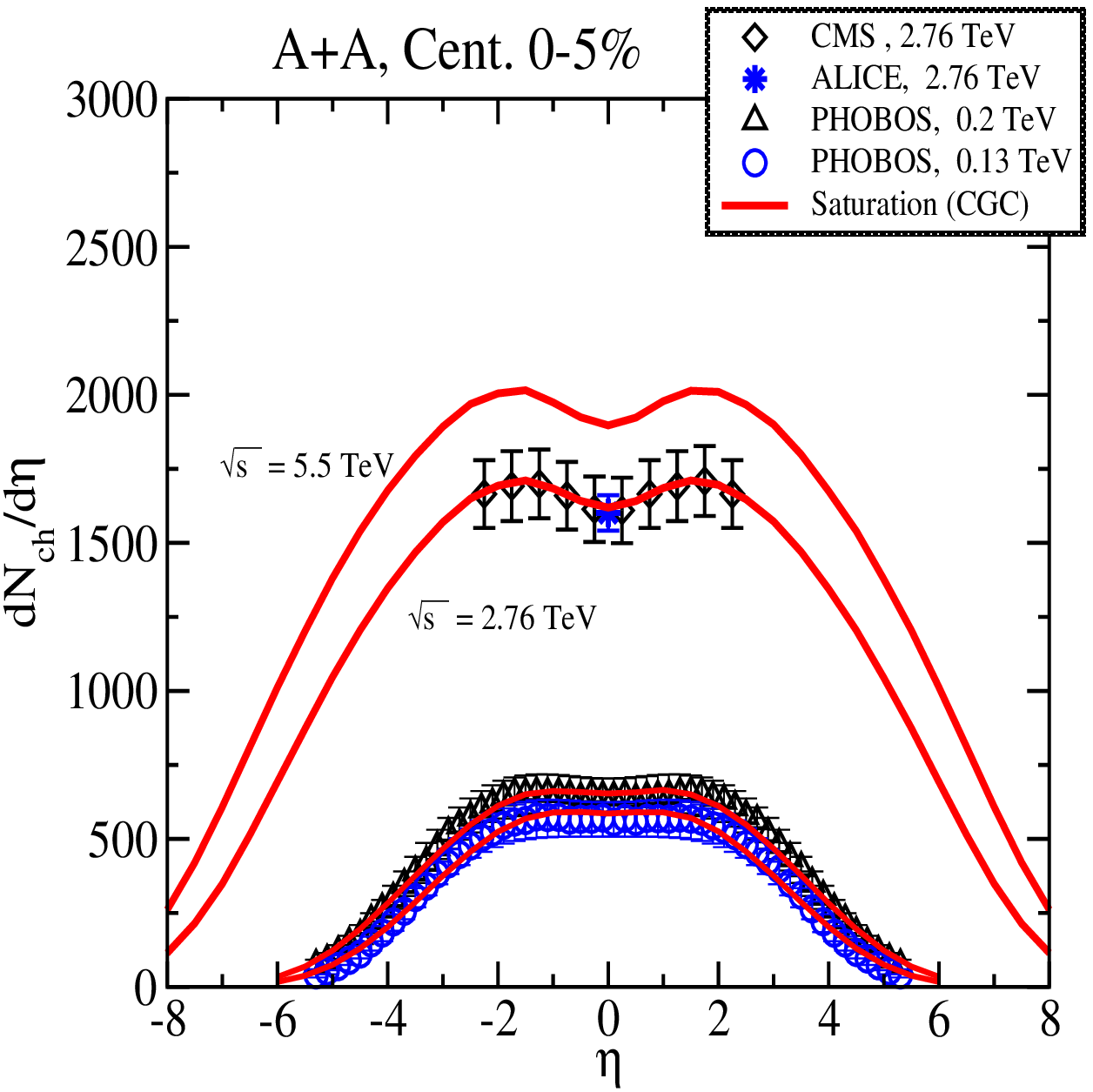}
                                  \includegraphics[width=8 cm]
                                  {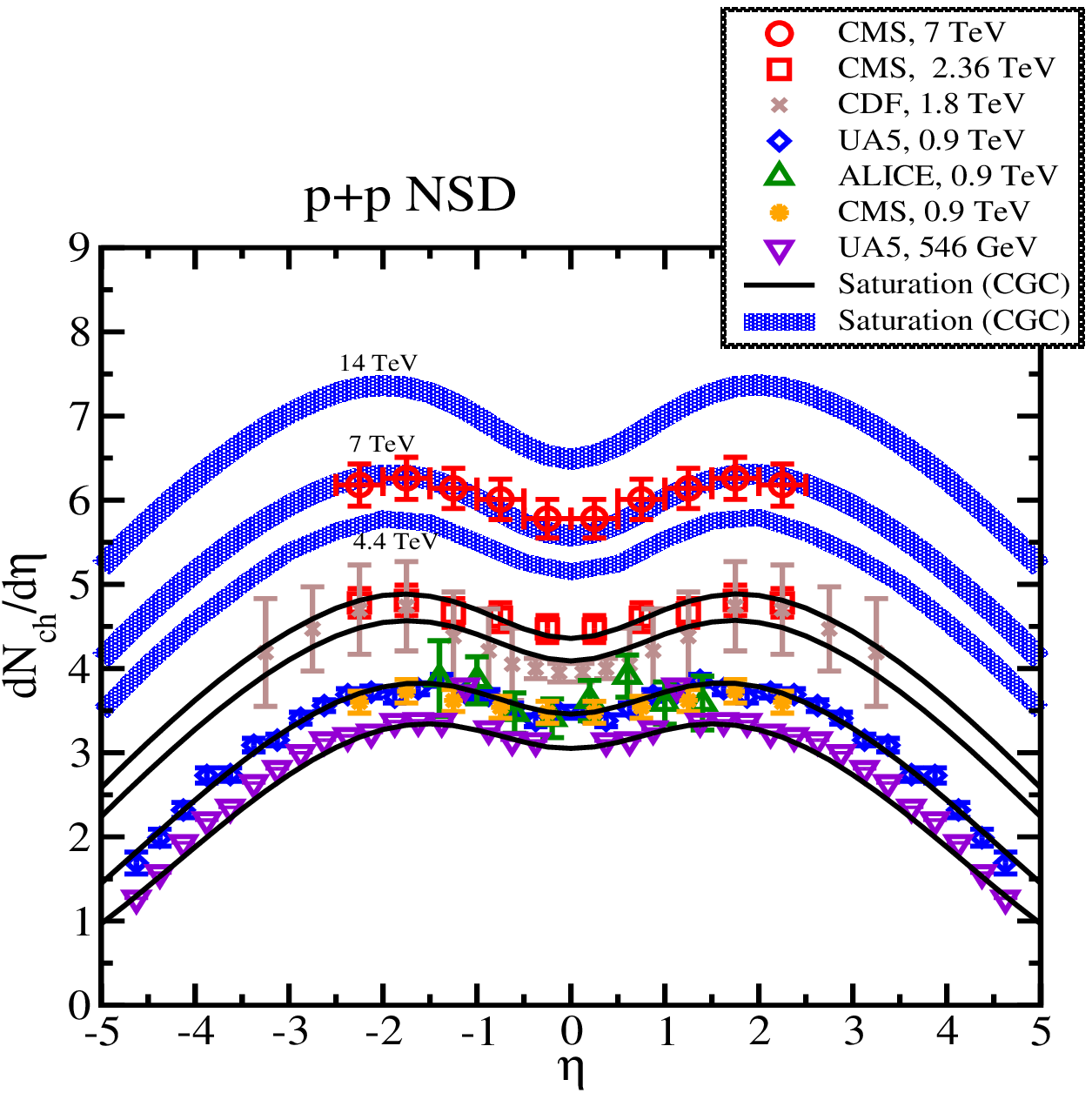} 
\caption{Pseudo-rapidity distribution of charged particles production
in p+p (right panel) and A+A (left panel) collisions at various energies. Right: the
 theoretical curves from top to bottom are at $\sqrt{s}=14, 7$ and $4.4$ TeV.  
The band in the right panel indicates about $2\%$ theoretical
error. The total theoretical uncertainties is less than $6\%$ mainly due to
our freedom in fixing the overall normalization and mini-jet mass with the experimental data at lower energy, see Ref.~\cite{ap}. The  experimental data are from \cite{CMS,Apb1,ppd}.  }
\label{fig1}
\end{figure}

First, in
\fig{fig1}, we show our description of the existing experimental data in p+p
inelastic non-singlet diffractive (NSD) collisions and predictions for
higher energies. Note that our curve at 7 TeV was prediction and it is seen that it is 
in good agreement with the recent LHC data \cite{ppd}. In \fig{fig1} (right), we also
show our predictions for charged hadron multiplicity distribution in
p+p collisions at $\sqrt{s}=4.4, 14$ TeV ($s$ is the center-of-mass
energy squared per nucleon).

In the case of scatterings on a nucleus target, one should also know
the atomic number (A) and impact-parameter dependence of the
saturation scale. In our approach, the saturation scale
of proton and nucleus both depend on the impact-parameter, see Refs.~\cite{me1,aa}. However, we
assume that the atomic number or $A$ (size of nuclei) dependence of
the saturation scale is approximately factorizable from energy or $x$
for a large nuclei \cite{me1,aa}. This assumption may be simply justified from the
experimental observation in A+A collisions that the centrality
dependence of the multiplicity at the LHC is very similar to RHIC up
to an overall normalization \cite{CMS}. Moreover, the pseudorapidity distribution of
multiplicity at a fixed energy but different centralities
approximately have similar shape upto a normalization
factor \cite{me1}. These experimental facts indicate that the energy and
centrality (or $A$) dependence of multiplicity, and as consequence the saturation scale,
can be approximately scaled out, in accordance with our assumption.  

As we already pointed out, one of the most interesting new feature of
the LHC multiplicity data is the fact that the energy growth of
multiplicities in A+A collisions is different from p+p ones
\cite{Apb1,CMS}. Notice that the $k_t$ factorization at the leading log
approximation accounts for most of hadron multiplicity while still
some contributions due to gluon decay cascade in the so-called MLLA
(Modified Leading Logarithmic Approximation) \cite{mlla} kinematics
may be missing \cite{me1}. The latter effect requires the inclusion of higher
order corrections and has been traditionally studied within a
different resummation scheme in which one systematically incorporates
next-to-leading logarithmic corrections, namely single and
double-logarithmic effects in the development of parton cascades
\cite{mlla}.  In Ref.~\cite{me1} we extracted
the energy-dependence of the gluon-jet decay cascade from $e^+e^-$
annihilation data and we showed that the energy-dependence of about
$s^{0.036}$ due to the enhanced gluon-decay cascade is exactly what
explains the different power-law energy-dependence of hadron
multiplicities in A+A compared to p+p collisions at the LHC. This effect
is more important for A+A collisions at high energy where the
saturation scale is larger and consequently the average transverse
momentum of the jet becomes about or bigger than $1$ GeV. The MLLA
gluon decay cascade gives rise to an extra contribution about
$20-25\%$ to the multiplicity in A+A collisions at the LHC. This
picture is fully consistent with all existing experimental data
including RHIC data \cite{me1}. We will show below that this picture
can be further tested in p+A run at the LHC.

In \fig{fig1} (left), we show the charged hadron pseudorapidity distribution
at the LHC in A+A central ($\sim 0-5\%$) collisions at various energies. The effect of the
gluon-decay cascade are incorporated in our results. Again, we have
only two parameters which was fixed with the lower energy data points at
the RHIC. Although, the midrapidity experimental data point at
$\sqrt{s}=2.76$ TeV for A+A $0-5\%$ central collisions was available \cite{Apb1} at
the time that we published our results \cite{me1}, the pseudorapidity
distribution was prediction and its agreement with the recently
released data from the CMS collaboration \cite{CMS} is striking.

\begin{figure}[t]                            
                                  \includegraphics[width=8 cm]
                                  {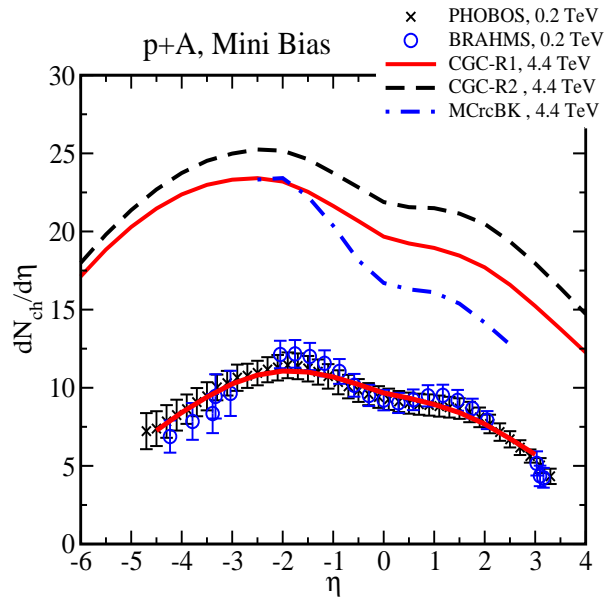} \caption{
                                  Pseudo-rapidity distribution of the
                                  charged particles production in p+A
                                  minimum (Mini) bias collisions at the LHC
                                  $\sqrt{s}=4.4$ TeV. The theoretical
                                  curves labeled by CGC-R1 and MCrcBK \cite{JA} are
                                  based on the leading log $k_t$-factorization formalism but
                                  two different saturation models. The
                                  curve labeled by CGC-R2 is based on the
                                  modified $k_t$-factorization formalism
                                  incorporating the gluon-decay
                                  cascade effects from the Modified Leading
Logarithmic Approximation scheme, see the text for the details. The experimental data are from \cite{pa-rhic}. }
\label{fig2}
\end{figure}

Having described both p+p and A+A multiplicities experimental data for
large range of energies/rapidities, the multiplicity in p+A collisions
can be considered as an important check of consistence.  The normalization again (similar to
\fig{fig1}) is fixed with RHIC data.  Unfortunately the error bars of RHIC data is rather large (see \fig{fig2}) and this induces uncertainties upto $\sim 15\%$ in our results.  It was shown in Refs.~\cite{me1,ap} that assuming 
a different mini-jet $m_{jet}$ for the cases of A+A and p+p collisions
will be in favor of data. Here, in order to reduce the model
dependence associated with the choice of mini-jet mass, we take a
fixed value for $m_{jet}$ for all rapidities, in the both
fragmentation region of nucleus and proton. We estimated that
uncertainties due to various value for $m_{jet}$ brings about $10\%$
errors in our calculation. We focus only on a pseudorapidity region where the effect of valence
quarks can be ignored. In \fig{fig2}, we show our prediction for the
pseudorapidity distribution of charged hadron multiplicity at the LHC
energy $\sqrt{s}=4.4$ TeV in minimum-bias p+Pb collisions. We show our
results both based on the $k_t$ factorization (labeled by CGC-R1) and
also the modified $k_t$ factorization incorporating the effect of the MLLA
gluon-decay cascade (labeled by CGC-R2). Notice that the contribution of the
MLLA gluon-decay cascade may be also important in p+A collisions at the LHC
energy since the average saturation scale of the system can be bigger
than $1$ GeV. This effect enhances the multiplicity about $20\%$,
in accordance with our results for A+A collisions \cite{me1}.

Notice that in the case of p+p and A+A collisions where the interacting system is
symmetric in the rapidity, the corresponding multiplicity distribution
is also symmetric (see \fig{fig1}). The position of two peaks in
pseudorapidity corresponds to the fragmentation region of the
projectile and the target. It has been shown that the peak in the
multiplicity distribution at forward and backward rapidity is enhanced
at higher energies due to the fact that the saturation scale increases
with energy (and density) \cite{me1,ap,aa,raj,kln,rs}. The exact shape of multiplicity
distribution (position of the peaks, the local minimum and the width
of the peak) depends on the unintegrated gluon density and the
Jacobian transformation which relates the rapidity and
pseudorapidity. In the case of p+A collisions, the interacting system
is not symmetric namely the saturation scale in the nuclear fragmentation
region is bigger than the projectile (proton) side and consequently
the multiplicity distribution in pseudorapidity is not symmetric.
In \fig{fig2}, we see rather strong forward-backward asymmetry for p+Pb compared to the
symmetric collisions. The shape and position of maximum in backward
and forward rapidity distribution are intricately related to the
saturation scale in nucleus and proton. A bigger saturation scale in
the nucleus washes away the maximum of the multiplicity distribution
in the proton fragmentation region.

For comparison, in \fig{fig2} we also show the predictions coming
from the MCrcBK of Ref.~\cite{JA}. Assuming that both predictions have about
$15\%$ errors due to the uncertainties associated with fixing the
normalization with the RHIC data, it is seen that two approaches give
rather different predictions for the multiplicity distribution in p+A
collisions. Using the $k_t$ factorization at the leading log
approximation, our prediction for charged hadron multiplicity at
$\sqrt{s}=4.4$ TeV at midrapidity $\eta=0$ in p+A mini-bias collisions
is $dN/d\eta=19.67\pm 1.5$ (CGC-R1) while prediction coming from the
MCrcBK approach \cite{JA} is $dN/d\eta=16.71\pm 1.3$ (MCrcBK). The main difference
between these two approaches is due to different employed saturation model.  On the
other hand, including the MLLA gluon decay cascade effect into the $k_t$ factorization increases our
multiplicity results and at midrapidity we have $dN/d\eta=21.88\pm 1.7$ (CGC-R2), see
\fig{fig2}. 

To conclude, in this letter, we provided quantitative predictions for
charged hadron multiplicity in p+Pb collisions at the LHC. We showed that the LHC p+Pb data on the
charged particles pseudo-rapidity multiplicity distribution can discriminate between saturation models and also
examine the $k_t$ factorization at an unprecedented level. In general,
the measurement of inclusive hadron production in p+A
collisions at the LHC can be a very good probe of the small-x physics, see also Ref.~\cite{pa}.

%---------------------------------------------------------------------
\begin{acknowledgments}
 The author would like to thank Jamal Jalilian-Marian and Genya Levin
 for useful discussion. The author is thankful to Adrian Dumitru for useful correspondence
and for providing the data points of the MCrcBK from Ref.~\cite{JA}. This
 work is supported in part by Fondecyt grants 1110781.
\end{acknowledgments}

%---------------------------------------------------------------------

\end{document}